\documentclass{esagnc}

\usepackage{booktabs}
\usepackage{siunitx}

\graphicspath{{./figures/}}

\newcommand{\RB}{R_{\mathrm{B}}}

\title{Continuous Learning of Gravity Field Irregularities Around
Small Bodies via Neural Hamiltonian ODEs}

\author[1,*]{Giacomo Acciarini}
\author[1]{Dario Izzo}

\affil[1]{Advanced Concepts Team, European Space Agency, ESTEC, Noordwijk, The Netherlands}
\affil[*]{Corresponding author: giacomo.acciarini@gmail.com}

\authorsInShort{G. Acciarini, D. Izzo}
\copyrgt{The Authors}

\begin{document}

\maketitle

\begin{abstract}
We propose to learn the unknown dynamics in the proximity of a small body directly from tracking data, representing them as a feed-forward neural network embedded in the system Hamiltonian. The equations of motion form a Neural Hamiltonian Ordinary Differential Equation, whose variational equations provide exact training gradients: estimation uses position and velocity arcs at realistic noise levels, without acceleration or potential labels, and a continual learning approach warm-starts the network as new data are acquired. The known part of the Hamiltonian carries whatever is available, from the central term and spin state to the constant-density model of the imaged shape. We assess the method against a normalized spherical harmonics expansion estimated from identical arcs through the same machinery, on scenarios built on the shapes of Itokawa, 67P, Bennu and Eros. The network remains usable inside the Brillouin sphere: it plans ballistic descents at Itokawa to \SI{4.6}{m} median touchdown error from tracking alone, against 5.1--\SI{48.9}{m} for harmonics of degree 4--12, and to \SI{0.9}{m} with the imaged shape as prior, a configuration that also recovers localised density anomalies invisible to any harmonics degree. The two representations are complementary, and we discuss their combined use across the phases of a small-body mission.
\end{abstract}

\section{Introduction}\label{sec:intro}

The gravity field of a small body is both a primary science product
and a core element of the force model used by the guidance,
navigation and control system during proximity operations. Its
estimation is subject to an important constraint related to the fact that a spacecraft does not directly measure gravity. Accelerometers sense only
non-gravitational forces, and the field enters the observables
exclusively through the trajectory, sampled as tracking data over
arcs that the navigation team is willing to fly. Any gravity model
intended for operational use must therefore work with noisy observed quantities (e.g.\ position
and velocity) and be estimated through the dynamics, from this kind
of data, at the level of noise present in it.

A well established representation for this task is the spherical
harmonics expansion, estimated by batch least squares from tracking,
from NEAR at Eros \cite{miller2002eros} to OSIRIS-REx at Bennu
\cite{scheeres2019bennu}. Its main structural limitation is equally
known \cite{takahashi2013surface}: the expansion converges only
outside the Brillouin sphere, the smallest sphere centered at the
origin of the expansion and containing the body, and by indicating with $R_B$ the Brillouin sphere radius and $r$ the radius at which the gravitational field is being evaluated, each coefficient
of degree $\ell$ is amplified by $(\RB/r)^{\ell}$ when the model is
evaluated below it. For elongated bodies the affected region is far
from marginal: as we quantify in Section~\ref{sec:deep}, it can contain a large portion of the close-proximity envelope of irregular bodies like Eros, Itokawa, and 67P.

In practice, this limitation is handled by switching gravity models
between mission phases: the harmonics expansion is used during the
orbital phase, where it converges and is estimated from the tracking,
and it is replaced, for descent and proximity operations, by a
polyhedral or mascon model computed from the imaged shape, often under an
assumed (typically uniform) interior density
\cite{werner1996polyhedron}. This procedure has two inherent limitations: first, the
proximity-operations model is not estimated from tracking data, so
these measurements never enter it and cannot automatically refine it further;
second, being derived from the imaged shape under an assumed
interior, its error is by construction largest exactly for the
gravitational signatures of internal irregularities.

Machine-learned gravity models have recently been proposed as an
alternative representation: neural density fields
\cite{izzo2022geodesy} represent the interior density of the body
with a network and obtain the potential by quadrature, while
physics-informed networks \cite{martin2022pinn,martin2022pinnsmall}
represent the potential directly, with the Laplace equation as a
regulariser. These works demonstrate that neural models can
represent the field of highly irregular bodies with considerable
accuracy; their training, however, relies on field labels, typically
$10^{4}$ to $10^{7}$ acceleration evaluations sampled at freely
chosen positions. This poses both the challenge of obtaining such
well-distributed data around the body, which is often not realistic
in mission operations, and the one of obtaining noiseless
gravitational measurements, which are in practice never directly
available. Whether, and where, a neural gravity model is useful in
the estimation setting of an actual mission, estimated from
tracking arcs alone, on flyable orbits and at realistic noise
levels, against the classical pipeline \cite{tapley2004statistical}
fitted to the very same data, and assimilating tracking as the
mission progresses, has to our knowledge not been established,
and is the subject of this paper.

\begin{figure}[t]
\centering
\includegraphics[width=0.70\linewidth]{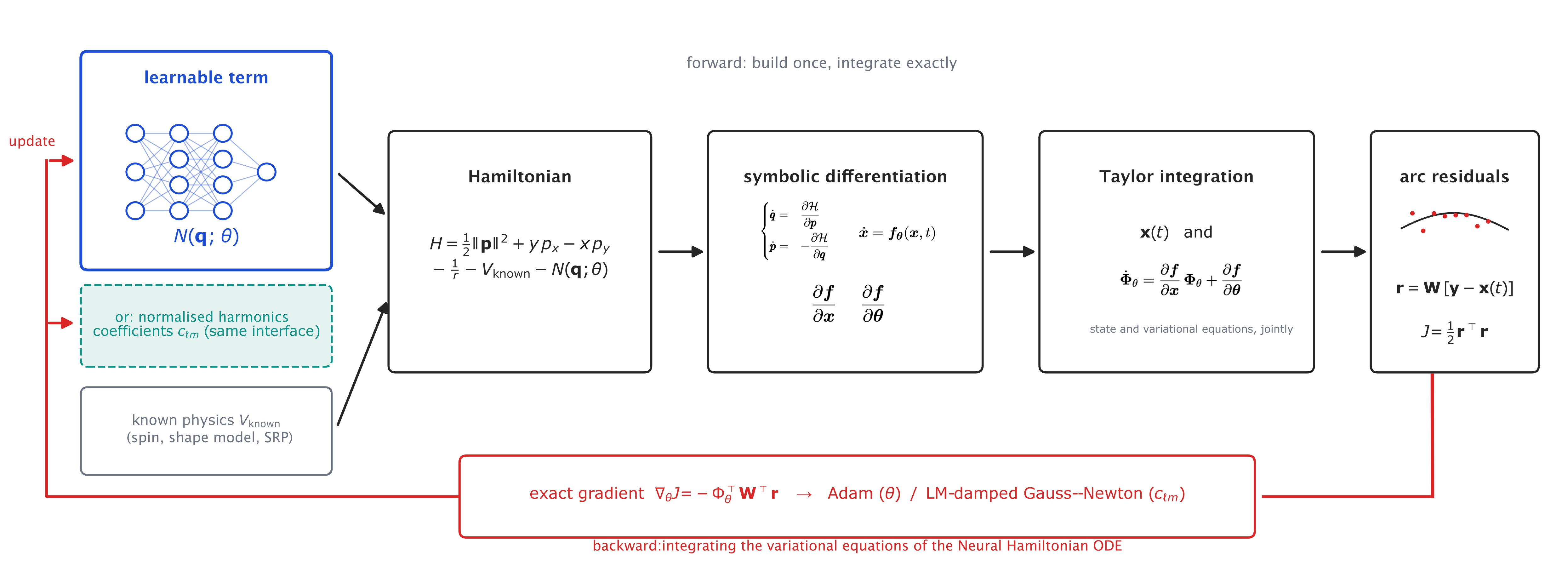}
\caption{The estimation pipeline. The learnable term (network weights
$\theta$, or harmonics coefficients $c_{\ell m}$ through the same
interface) enters the Hamiltonian of Eq.~\eqref{eq:ham}; Hamilton's
equations of motion are derived and the variational equations are
constructed symbolically; the augmented system is then integrated
forward, providing the predicted arcs together with their exact
sensitivities to the parameters, from which the residual and its
gradient are assembled for the update.}
\label{fig:schematic_illustration}
\end{figure}

We formulate the problem as the estimation of the unknown part of
the system Hamiltonian. The known part contains whatever is
available at the time of the fit, which follows the timeline of a
mission: initially only the central term and the spin state; later,
the constant-density model of the imaged shape; throughout, any
measured non-gravitational forces. The learned part is a small
feed-forward network, compiled symbolically into the equations of
motion from the Hamiltonian, and integrated forward in time via a Taylor integrator~\cite{biscani2021heyoka}. Due to the neural term in the Hamiltonian, the equations of motion form a Neural Hamiltonian Ordinary Differential Equation (ODE)~\cite{greydanus2019hamiltonian}, and the variational equations of the flow with respect to the network's parameter provide the evolution of the exact gradients that can be used for training. Once the neural term is learned, a high-order expansion of the
neural Hamiltonian flow can also be leveraged to map uncertainties
at future times through the nonlinear dynamics, as recently
demonstrated on the Didymos--Dimorphos binary, the target of ESA's
Hera mission~\cite{acciarini2025nonlinear, izzo2025high,
michel2022hera}. The same machinery is also used for the baseline model: a normalized
spherical harmonics expansion, whose coefficients enter the
integrator as runtime parameters exactly as the network weights and biases do.
Every comparison in this paper is therefore between representations
estimated from identical data through identical workflows: the
harmonics baseline is never taken from the literature but
re-estimated in every experiment, through the same flow, with the
same exact Jacobians, and with regularisation where it helps. Our proposed
formulation belongs to the family of Hamiltonian and neural-ODE
models of learned dynamics
\cite{chen2018neuralode,greydanus2019hamiltonian}, which has
recently shown promising results in astrodynamics applications and
beyond~\cite{acciarini_space, izzo2025high}.
The schematic of the proposed workflow and the baseline (via the spherical harmonics coefficients) is shown in Fig.~\ref{fig:schematic_illustration}.

The main contributions are the following. (i) We show that, from
tracking alone and with no shape model, the neural Hamiltonian approach
provides a gravity model that can consume tracking data during descent and close-proximity operations, which is valid inside and outside
the Brillouin sphere and that in descent scenarios it
outperforms the harmonics fit. (ii) We show that the
formulation extends naturally to the joint estimation of
non-gravitational Hamiltonian forces, and that it is more robust to
force model errors than the harmonics fit.
(iii) We show that, once the imaged shape is available, the correct
architecture is a residual one, with the shape model in the known
part of the Hamiltonian, and that in this configuration the network
can recover localized interior structures that are not observable via spherical harmonics of any degree. (iv) We discuss a generic continual learning procedure that can be used to train neural Hamiltonian networks in the context of geodesy, as new data is being acquired during the mission. 
\section{Methodology}\label{sec:method}

\subsection{Dynamics}

We work in the body-fixed rotating frame, in the nondimensional
canonical coordinates defined by the body's gravitational parameter
and spin period. For a single spinning body the Hamiltonian is
\begin{equation}
H(\bm q,\bm p;\bm\theta)
 = \tfrac12\lVert\bm p\rVert^{2} + y\,p_x - x\,p_y
 - \frac{1}{r}
 - V_{\mathrm{known}}(\bm q)
 - \mathcal{N}(\bm q;\bm\theta),
\label{eq:ham}
\end{equation}
where $V_{\mathrm{known}}$ contains the terms known at the time of
the fit (nothing, in the experiments of Section~\ref{sec:noshape};
the perturbing potential of the constant-density shape model, and
where applicable a modelled solar radiation pressure term, in
Section~\ref{sec:shape}), and $\mathcal{N}$ is the estimated correction. The
canonical momenta are related to the rotating-frame velocity by
$v_x = p_x + y$, $v_y = p_y - x$, $v_z = p_z$. Lengths and times are
normalized so that the gravitational parameter and the frame rotation
rate are unity: by the synchronous radius of the spinning body and
its spin period over $2\pi$.

The ground truth gravitational field is constructed via mascon models \cite{muller1968mascons}.
In the truth, the body's own gravity (the central term, the
shape-model potential and the correction $\mathcal{N}$ of Eq.~\eqref{eq:ham}
taken together) is replaced by the full mascon sum, while any
physics external to the body's gravity is common to the truth and to
the fitted model:
\begin{equation}
\mathcal{H}_{\rm truth} = \tfrac12\lVert\bm p\rVert^{2}
 + y\,p_x - x\,p_y
 - \sum_{i=1}^{N_m} \frac{\mu_{m,i}}{r_{m,i}} - V_{\rm known},
\label{eq:truth}
\end{equation}
where $\mu_{m,i}$ are the normalized mascon masses, summing to the
body's normalized gravitational parameter, $r_{m,i}$ are their
distances from the spacecraft, and $V_{\rm known}$ collects the
external terms where present, i.e.\ the solar radiation pressure
term in the scenarios where SRP is active. The mascon sets are
synthetic bodies, or the
high-resolution mascon models of Itokawa, 67P, Bennu and Eros, in some experiments modified
with heterogeneous interior density; all are propagated with the same
Taylor scheme, and the estimators never see
anything but the resulting trajectories.

Two candidate models are used for the estimated correction. The neural candidate is a feed-forward network
$\mathcal{N}_{\bm\theta}$ with up to two hidden layers of 16 or 24
units and $\tanh$ activations, which results in 353--721 learnable parameters $\bm\theta$ in the
experiments, mapping the body-fixed position to a scalar potential
correction and multiplied by a fixed radial envelope,
\begin{equation}
\mathcal{N}(\bm q; \bm \theta) = s \, w(r_1)\, \mathcal{N}_{\bm\theta}(x,y,z),
\qquad
w(r_1) = \left(\frac{R_w}{r_1}\right)^{3},
\label{eq:neural}
\end{equation}
with $r_1$ the distance from the body centre, and $R_w = 1$ the nondimensional envelope radius.
The $w$ factor in front of the network is
there for simple reasons. From far away every body looks like a
point mass, and the point mass is already in the known term; what is
left for the network is the signature of the irregular shape and
interior, which fades with altitude, and its slowest-fading part
falls off as $1/r_1^{3}$ in the potential. The
factor $w(r_1)$ imposes exactly this falloff: near the body the
correction is free to be rich, while far from it the learned term
fades at the physical rate whatever the learnable parameters do.
The scale $s$ can be used to bring the output of a freshly initialized
network to the typical size of the unmodelled potential in these
units, so that training starts in the right range: in our experiments we observe that $s=0.1$ seems to be a good choice. Nothing else is imposed: in particular
$\mathcal{N}_{\bm\theta}$ is not constrained to be harmonic, which
is what lets the same object remain valid on both sides of the
Brillouin sphere. In the experiments in
Sections~\ref{sec:noshape} and~\ref{sec:shape}, we will observe what that freedom costs and buys above and below the Brillouin sphere.
One subtlety of placing a network inside a Hamiltonian is
that since $\mathcal{H}$ is defined up to an additive constant, a
position-only correction with a linear output layer and no envelope
would leave the bias of the last layer with identically zero
gradient, but the $w$ factor removes this degeneracy here.

The baseline correction is instead the normalized spherical harmonics
expansion of the perturbing potential,
\begin{equation}
\mathcal{N}_{\rm SH}(\bm q;\bm c) = \frac{\mu_1}{R}
\sum_{\ell=2}^{L}\sum_{m=0}^{\ell}
\left(\frac{R}{r_1}\right)^{\ell+1}
\bar P_{\ell m}(\sin\varphi)
\left[\bar C_{\ell m}\cos m\lambda + \bar S_{\ell m}\sin m\lambda\right],
\label{eq:sh}
\end{equation}
where $(r_1,\varphi,\lambda)$ are body-centred radius, latitude and
longitude, $R$ is the reference radius of the expansion (set to the
Brillouin radius $\RB$), $\mu_1$ is the body's nondimensional
gravitational parameter, and $\bar P_{\ell m}$ are the fully
normalized associated Legendre functions. The estimated parameter
vector $\bm c$ collects the normalized coefficients
$\bar C_{\ell m}$ and $\bar S_{\ell m}$ for $\ell = 2,\dots,L$: the
degree-0 term is the known monopole, and the degree-1 terms vanish when writing the frame with origin
in the centre-of-mass, leaving $(L{+}1)^{2}-4$ free
coefficients (21, 77, 165 and 285 at $L = 4, 8, 12, 16$). The
expansion and its derivatives are evaluated in Cartesian coordinates
with the Cunningham recursions \cite{cunningham1970}, which are free
of polar singularities.
\subsection{Backpropagating through the Flow}
\label{sec:estimation}

\begin{figure}[t]
\centering
\includegraphics[width=0.80\linewidth]{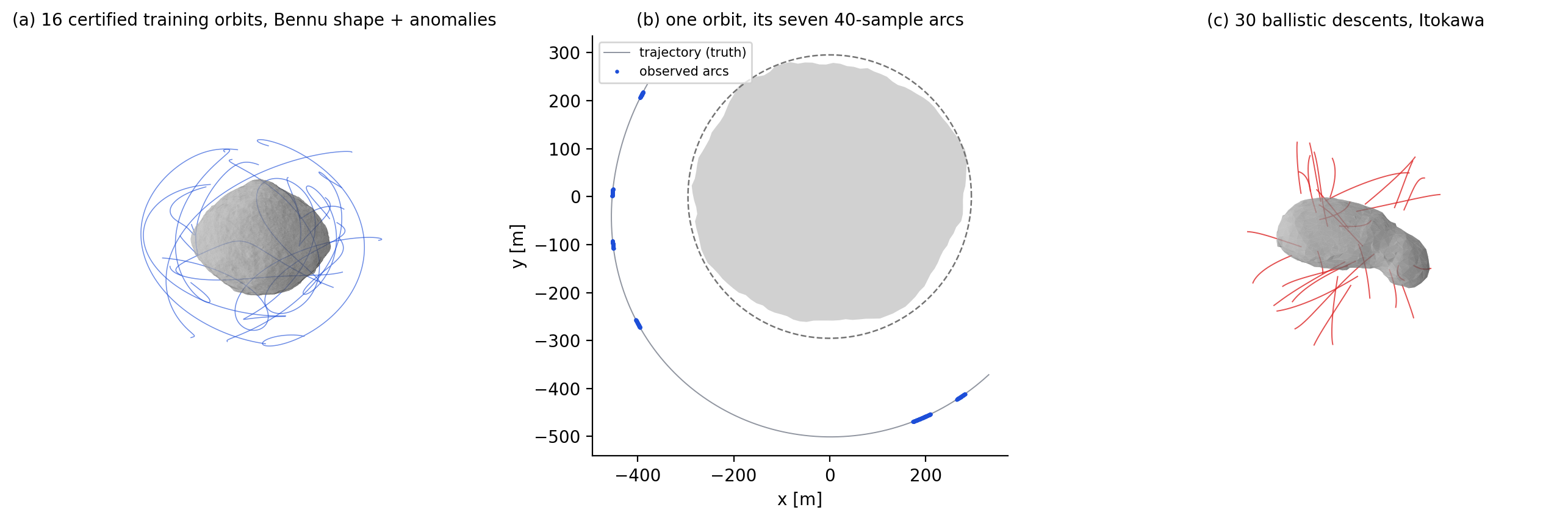}
\caption{Trajectories and observables of the experimental campaigns.
(a) The sixteen training orbits around the Bennu-shaped body with
buried anomalies of Section~\ref{sec:nuggets}. (b) One of the orbits at Bennu with its seven observed arcs of
40 samples each; the filled silhouette is the body, the dashed
circle the Brillouin sphere. (c) The thirty ballistic descents to
Itokawa of Section~\ref{sec:touchdown}.}
\label{fig:orbits}
\end{figure}

Hamilton's equations applied to Eq.~\eqref{eq:ham} yield the
equations of motion $\dot{\bm x} = \bm f(\bm x;\bm\theta)$ for the
state $\bm x = (\bm q, \bm p)$. The observables are
position--velocity arcs: windows of $m=40$ samples $\bm y_{kj}$ at
times $t_{kj}$, indicating the $j$-th (noisy) element (with $j=0,\dots, m-1$) of the $k$-th arc. Each sample within an observation arc is taken at a 14 seconds cadence, so that each observation arc lasts roughly 10 minutes. Uncorrelated Gaussian noise is applied to the
positions ($\sigma_r = \SI{5}{cm}$) and to the rotating-frame
velocities ($\sigma_v = \omega\,\sigma_r/10$, i.e.\ one tenth of the
position noise per nondimensional time unit, corresponding to
0.7--2.0\,\si{\micro m\,s^{-1}} across the bodies considered), and
the noisy velocities are mapped to the canonical momenta through the
linear relations above. All training
trajectories are non-impacting ones. Fig.~\ref{fig:orbits} shows a
training set, one orbit with its arcs, and the descent campaign of
Section~\ref{sec:touchdown}; Table~\ref{tab:campaigns} summarises
the observation content of every experiment.

\begin{table}[t]
\centering\footnotesize
\caption{Observational campaigns of the paper. ``Scalars'' is the total number of
scalar measurements ($\text{arcs}\times 40 \times 6$, hence including the fact that the state is 6-dimensional). ``net/SH'' gives the size of the estimated
parameter vector for the network and for the harmonics degrees used
in that experiment; ``prior'' indicates whether a mascon model with
constant-density of the imaged shape is carried in
$V_{\mathrm{known}}$.}
\label{tab:campaigns}
\setlength{\tabcolsep}{2.3pt}
\begin{tabular}{lllrrlc}
\toprule
experiment & truth body & orbits & arcs & scalars & net / SH & prior\\
\midrule
\ref{sec:deep} deep region & 67P homog. & 18, 6 stages (+9 desc.) & 126 (+9) & 32.4k & 721 / 77--165 & no\\
\ref{sec:touchdown} descent planning & Itokawa 2$\times$ & 16, 1.2--2.0\,$\RB$, 10--80$^{\circ}$ & 112 & 26.9k & 721 / 21--165 & no\\
\ref{sec:touchdown} extended campaign & Itokawa 2$\times$ & 48, 1.2--2.0\,$\RB$, 10--80$^{\circ}$ & 672 & 161.3k & 721 / 21--77 & no\\
\ref{sec:srp} particle swarm & Bennu & 28 particles & 263 & 63.1k & 353+28 / 49--193 & no\\
\ref{sec:architecture} interiors & Itokawa/hollow heter. & 16 & 112 & 26.9k & 721 / 21--165 & yes\\
\ref{sec:resid-touchdown} descent, residual & Itokawa 2$\times$ & 16, 1.2--2.0\,$\RB$ & 112 & 26.9k & 721 / 21--77 & yes\\
\ref{sec:nuggets} anomalies & Bennu + 4 anom. & 16, 1.08--1.7\,$\RB$, 10--80$^{\circ}$ & 112 & 26.9k & 721 / 21--285 & yes\\
\bottomrule
\end{tabular}
\end{table}

The estimation problem is the batch least squares of orbit
determination practice \cite{tapley2004statistical},
\begin{equation}
\hat{\bm\theta} = \arg\min_{\bm\theta} J(\bm\theta), \qquad
J(\bm\theta) = \tfrac12\,\bm r^{\top}\bm r, \qquad
\bm r_{kj}(\bm\theta) = \bm y_{kj}
 - \bm x(t_{kj};\bm x_{k0},\bm\theta).
\label{eq:cost}
\end{equation}
In this case, however, each arc is anchored at its first observed sample $\bm x_{k,0}=\bm y_{k,0}$, rather than an estimated epoch state.

The residuals are formed directly on the nondimensional canonical
states. The design matrix requires the state
sensitivities, which are obtained from the first-order variational
equations, integrated jointly with the state:
\begin{equation}
\dot{\bm\Phi}_{\theta}
 = \frac{\partial \bm f}{\partial \bm x}\,\bm\Phi_{\theta}
 + \frac{\partial \bm f}{\partial \bm\theta},
\qquad
\bm\Phi_{\theta}(t_0) = \bm 0,
\qquad
\bm\Phi_{\theta}(t) = \frac{\partial \bm x(t)}{\partial \bm\theta}.
\label{eq:var}
\end{equation}
Since the network (and, equally, the harmonics expansion) is compiled
symbolically into $\bm f$, the Jacobians appearing in
Eq.~\eqref{eq:var} are closed-form expressions, and the Taylor
integrator propagates Eq.~\eqref{eq:var} at the same tolerance as the state.
Stacking the residuals of Eq.~\eqref{eq:cost} over all $M$ samples
of all arcs into a single vector $\bm r \in \mathbb{R}^{6M}$
(Table~\ref{tab:campaigns} reports $6M$ as ``scalars''), its exact
Jacobian with respect to the $n_p$ estimated
parameters is the design matrix
\begin{equation}
\bm A \;=\; \frac{\partial \bm r}{\partial \bm\theta}
 \;=\; -\begin{bmatrix}
 \bm\Phi_{\theta}(t_{11})\\
 \bm\Phi_{\theta}(t_{12})\\
 \vdots
 \end{bmatrix}
 \;\in\; \mathbb{R}^{6M\times n_p},
\label{eq:design}
\end{equation}
one $6\times n_p$ block per sample $(k,j)$. The cost gradient
$\nabla_{\theta} J = \bm A^{\top}\bm r$ is therefore computed exactly via the variational equations.

\subsection{Solvers}\label{sec:solvers}

The harmonics' coefficients $\bm c$ enter the dynamics nearly
linearly (since the tracking arcs are relatively short), and Eq.~\eqref{eq:cost} is solved for them by Gauss--Newton
iterations with Levenberg--Marquardt damping~\cite{fletcher1971modified},
\begin{equation}
\left(\bm A^{\top}\bm A
 + \lambda\,\mathrm{diag}(\bm A^{\top}\bm A)\right)\delta\bm c
 = -\bm A^{\top}\bm r,
\label{eq:lm}
\end{equation}
where each step is accepted if the cost decreases
($\lambda \to \lambda/3$, floored at $10^{-14}$) and rejected
otherwise ($\lambda \to 10\lambda$), starting from
$\lambda = 10^{-6}$ and $\bm c = \bm 0$. Convergence to a relative
cost change below $10^{-6}$ typically requires three iterations.

The high degrees of the expansion are weakly observable from orbit:
the same $(\RB/r)^{\ell}$ attenuation that amplifies a coefficient
below the Brillouin sphere suppresses its signature in arcs flown
above it. This means that high-degree coefficients are nearly unconstrained and are free to absorb noise far away from the surface, and this is then amplified near the surface with larger errors. The standard way to prevent this in gravity estimation practice is a
Kaula-type constraint. Kaula's rule \cite{kaula1966} is the
empirical observation that planetary coefficient spectra decay with
degree as a power law, $\bar C_{\ell m}, \bar S_{\ell m} \sim
\ell^{-2}$ up to a body-dependent constant.

This transforms the equation above into
$(\bm A^{\top}\bm A + \alpha\bm K)\,\delta\bm c
 = -(\bm A^{\top}\bm r + \alpha\bm K\bm c)$ with
$\bm K = \mathrm{diag}(\ell^{4})$. Wherever the harmonics baseline
could benefit from this constraint, it is granted it: in
Section~\ref{sec:nuggets} the weight $\alpha$ is scanned over
$10^{-8}$--$10^{-4}$ and the value performing best on the
truth-based evaluation metric is retained: this means that the baseline is quite strong and also accounts for these effects known in standard geodesy practice.

For the case of the network's learnable parameters, their relationship with the cost is highly nonlinear, which makes
Eq.~\eqref{eq:lm} not appropriate for them. Instead, Eq.~\eqref{eq:cost} is minimised with Adam \cite{kingma2015adam} on the exact
gradient, mini-batched over arcs (batch size 16--32), with a cosine
learning rate schedule \cite{loshchilov2017sgdr}, and 10--15\% of the arcs held out for validation.

Both solvers leverage the $\bm A$ matrix computed in the same way (via exact variational equations integration), so that the
comparisons of Sections~\ref{sec:noshape} and~\ref{sec:shape} are comparisons between different representations and do not contain different Jacobian information.

\subsection{Continual Learning Approach}\label{sec:continual}

The aim of the proposed work is also its deployment/use onboard. For this to happen, one must account for the fact that tracking does not necessarily arrive immediately as one single batch, but more batches are collected as the mission evolves. Hence, we would like to avoid to refit from scratch the model for every batch, and instead warm-start it from previous batches.
The way this is done is that data arrives in different stages: at stage $k$, for instance we would have $\mathcal{D}_{1:k} = \mathcal{D}_1 \cup \dots \cup \mathcal{D}_k$, the cost in Eq.~\eqref{eq:cost} is minimized over the accumulated dataset, by warm starting the optimization from previous solutions (in case available):
\begin{equation}
\bm\theta_k = \arg\min_{\bm\theta}
J_{\mathcal{D}_{1:k}}(\bm\theta)
\quad \text{initialised at } \bm\theta_{k-1},
\label{eq:continual}
\end{equation}
where the iterate with the best validation loss is retained. The
harmonics baseline is treated identically, with the Gauss--Newton
iteration of Eq.~\eqref{eq:lm} warm-started at the previous
coefficients.

The same variational machinery also turns the arrival of data into a
decision variable. The sensitivities that drive the solvers also
measure how much a set of arcs constrains the parameters: stacking
Eq.~\eqref{eq:var} over the samples of the arcs into the design
matrix $\bm A$ of Section~\ref{sec:estimation} gives the
Gauss--Newton information matrix of the cost of Eq.~\eqref{eq:cost},
\begin{equation}
\bm\Lambda = \bm A^{\top}\bm A
 = \sum_{k,j}
 \bm\Phi_{\theta}(t_{kj})^{\top}\,\bm\Phi_{\theta}(t_{kj}),
\label{eq:info}
\end{equation}
an $n_p\times n_p$ matrix over the $n_p$ estimated parameters, whose
$(a,b)$ entry accumulates, over the sample times and over all six
canonical components, the products of the trajectory's sensitivities
to parameters $a$ and $b$. In the linear approximation
$\bm\Lambda^{-1}$ is proportional to the parameter covariance, so an
increase of $\log\det\bm\Lambda$ is a shrinkage of the parameter
confidence ellipsoid. Writing $\bm\Lambda_{\rm acc}$ for the
information accumulated by the orbits already selected, a candidate
orbit $c$ is then scored, before flying it, by the D-optimal
gain~\cite{pukelsheim1993optimal}
\begin{equation}
\Delta(c) =
\log\det\!\left(\bm\Lambda_{\rm acc} + \bm\Lambda_c
 + \varepsilon\bm I\right)
- \log\det\!\left(\bm\Lambda_{\rm acc} + \varepsilon\bm I\right),
\qquad
\varepsilon = \max\!\left(10^{-8},\,
10^{-6}\,\tfrac{\mathrm{tr}\,\bm\Lambda_{\rm acc}}{n_p}\right),
\label{eq:dopt}
\end{equation}
where both matrices are instances of Eq.~\eqref{eq:info} built from
predicted sensitivities alone, no measurements enter them, which
is what makes the score computable before any data exist. In simple terms, the criterion favours the candidate whose future arcs would most increase the curvature of the cost in the parameter directions that the arcs already flown leave flattest.
$\bm\Lambda_c$ is the information predicted for the candidate's
future arcs under the current model $\bm\theta_k$: the
candidate trajectory is propagated together with Eq.~\eqref{eq:var}
at $\bm\theta_k$. $\bm\Lambda_{\rm acc}$ is the
running sum of these predicted contributions over the orbits
selected so far; it starts from zero, which is one of the main
reasons for having $\varepsilon\bm I$: the score must remain
finite at the first selections, when $\bm\Lambda_{\rm acc}$ is
singular, and a network generically carries locally flat parameter
directions of near-zero information. The ground truth is
never consulted, so $\Delta(c)$ is a score a mission could compute
on board.  At each stage one can check a pool of candidate orbit families and greedily select maximizers of
$\Delta(c)$, updating $\bm\Lambda_{\rm acc}$ after every selection.

\begin{figure}[t]
\centering
\includegraphics[width=0.74\linewidth]{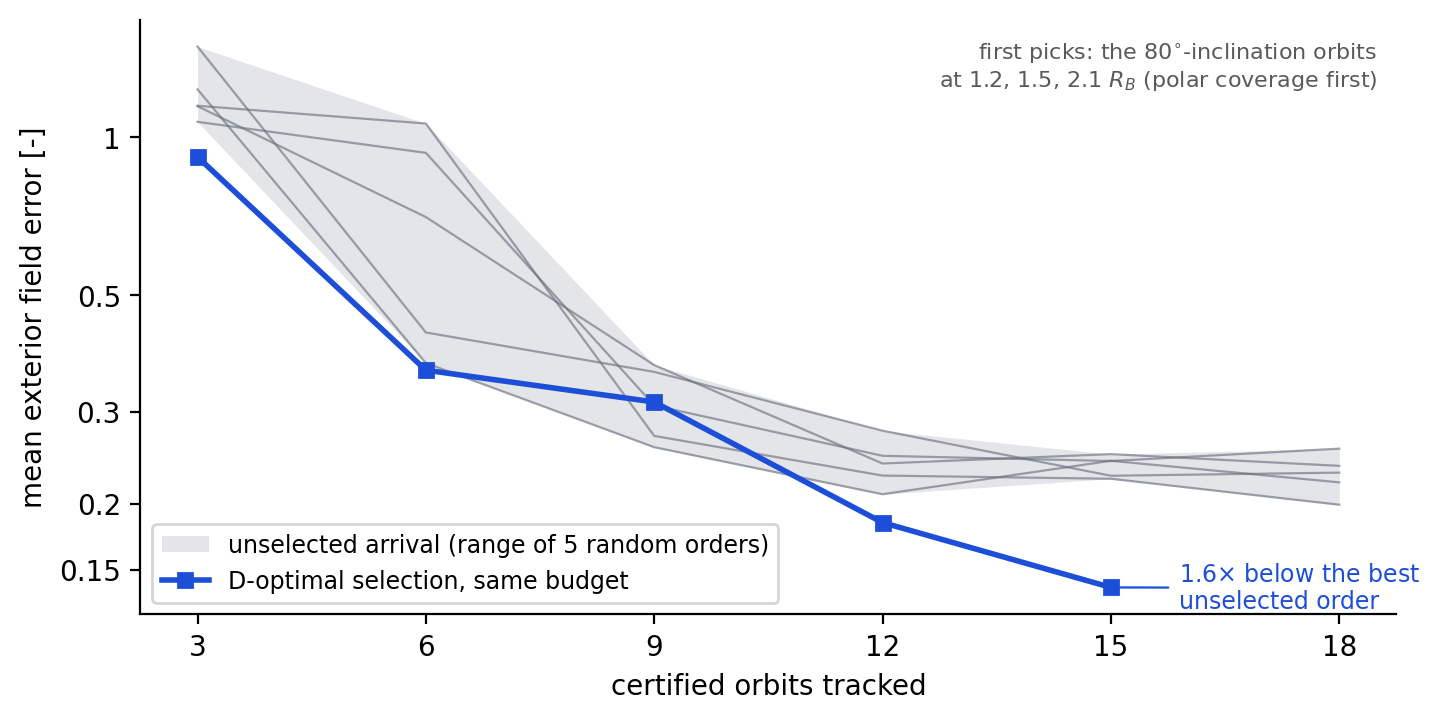}
\caption{Information-driven versus unselected data acquisition on
the 67P campaign of Section~\ref{sec:deep}: mean exterior field
error (shells 1.2--1.7\,$\RB$) versus number of orbits
tracked. Thin grey curves and shaded band: the warm-refitted network
under five random arrival orders of the same orbits. Bold:
the D-optimal selection of Eq.~\eqref{eq:dopt} on the same budget.}
\label{fig:continual}
\end{figure}

Fig.~\ref{fig:continual} shows the procedure at work on the 67P
campaign of Section~\ref{sec:deep}. The same campaign was run under five random arrival orders:
identical orbits, identical arcs and budgets, only the
permutation assigning orbits to stages changes. Chance matters,
after six orbits the luckiest order stands at 0.37 and the
unluckiest at 1.06, a factor of three apart, and eventually they settle between 0.20 and
0.26. The D-optimal campaign removes the lottery: it tracks the
luckiest orders through the early stages, is below all five from
twelve orbits onwards, and with fifteen orbits reaches 0.14,
$1.6\times$ below the best random order at equal budget, and better
than any unselected campaign achieves even with the full eighteen.
Its choices are also readable: the first picks are the three
$80^{\circ}$-inclination orbits across the radius range, polar
coverage before anything else.

\section{Results From Tracking Alone}\label{sec:noshape}

In all experiments of this section no shape model is used: the
fitted dynamics consist of the central term, the spin, and the
estimated correction $\mathcal{N}$. This corresponds to the situation of
approach and early proximity operations.

This section tackles the first four rows of
Table~\ref{tab:campaigns}: the deep-region campaign on 67P
(Section~\ref{sec:deep}), the ballistic-descent campaigns on Itokawa
at two data budgets (Section~\ref{sec:touchdown}), and the joint
gravity--SRP estimation on the Bennu particles
(Section~\ref{sec:srp}). Each subsection opens by defining the scenario and observation campaign, including truth model, data, estimators, and then
reports the comparison.

The estimators are trained on
position--velocity arcs alone and never see a field quantity;
evaluation, however, is a different matter. Given that this is a
simulation study, the
true field is available to us, though never to the estimators, and it can be used to evaluate the fitted models over a whole region,
including places where no observation arc is available, and independent of any particular trajectory or time horizon.
The accuracy
of a fitted field is reported as the
\emph{relative field error}
\begin{equation}
\epsilon \;=\;
\sqrt{\frac{\sum_{i=1}^{N_e}
       \left(V_{\rm model}(\bm q_i) - V_{\rm truth}(\bm q_i)\right)^{2}}
      {\sum_{i=1}^{N_e} V_{\rm truth}(\bm q_i)^{2}}},
\label{eq:fielderr}
\end{equation}
where $V$ denotes the perturbing potential (the target of the
correction) and $\{\bm q_i\}_{i=1}^{N_e}$ is an evaluation set of
positions: $\epsilon = 0$ is a perfect recovery and $\epsilon = 1$
is the accuracy of applying no correction at all. We use two kinds of evaluation sets: \emph{shells} are a few hundred random
directions at a fixed barycentric radius, quoted in units of $\RB$
(e.g.\ $1.2\,\RB$). Since at fixed $r/\RB$ the harmonics amplification
$(\RB/r)^{\ell}$ is uniform, the shells seem to be a very natural way to compare alternative models to the baseline. Then, we also employ \emph{surface-following sets}, quoted as
an altitude $h$ (e.g.\ $h = \SI{30}{m}$): these are the mesh vertices of
the shape displaced radially outward by $h$. On an irregular body
constant altitude is very different from constant radius (the
$h=\SI{30}{m}$ set on Itokawa spans $r = 0.36$--$1.11\,\RB$), and
these sets are useful to test the terminal-descent regimes. The altitudes are scaled to the body: \SI{30}{m}
for the sub-kilometre bodies,
\SI{300}{m} for the 4.6-km 67P.

Comparisons between an iteratively trained network and a
Gauss--Newton fit based on single runs are vulnerable to chance: the
two largest campaigns were therefore repeated eight times end to
end, the 67P experiment over redrawn noise and trajectory geometry
and the Bennu swarm over complete redraws of the particle ejections,
and results quoted below as ``$n$ of 8'' refer to these repetitions.
The central comparative statements hold in every repetition unless
noted.

\subsection{Inside the Brillouin Sphere: the 67P Scenario}
\label{sec:deep}

We first quantify how much of the operationally relevant volume is
affected by the divergence of the exterior expansion. As a proxy for
the volume in which close-proximity operations take place (hovering,
descent and ascent, sampling), we define the operational
envelope of a body as the layer of space between its surface and an
altitude of $0.15\,\RB$ above it, and we sample it geometrically,
from the shape model alone: every surface point is raised radially
to a set of altitudes up to $0.15\,\RB$, giving a cloud of positions
that covers the envelope. The question is then simple: what fraction
of these positions lies inside the Brillouin sphere, i.e.\ at a
barycentric radius $r < \RB$, where the exterior expansion diverges?
The answer is 96\% for Eros, 97\% for Itokawa, 93\% for 67P and 84\%
for Bennu: for an elongated body, essentially the whole
close-proximity volume sits where the classical representation is
invalid. In the
following we refer to the free space inside the Brillouin sphere as
the ``deep region''. The network model has no convergence boundary
and can be evaluated anywhere; whether training on exterior arcs
makes it accurate in the deep region is the question addressed by
this experiment.

The deep region is largest, among the bodies considered, on the
bilobed 67P, which is therefore the body chosen for this experiment. The signal to be
estimated can be easily assessed: relative to the central body attraction, the
unmodelled perturbing acceleration has a weight of 16\%, taken as a median over the positions at $1.1\, \RB$, and of 5\% at $2\,\RB$, and is still
above 2\% at $3\,\RB$. This is the signal
the estimators need to provide (i.e., through the term $\mathcal{N}$).

\begin{figure}[t]
\centering
\includegraphics[width=0.78\linewidth]{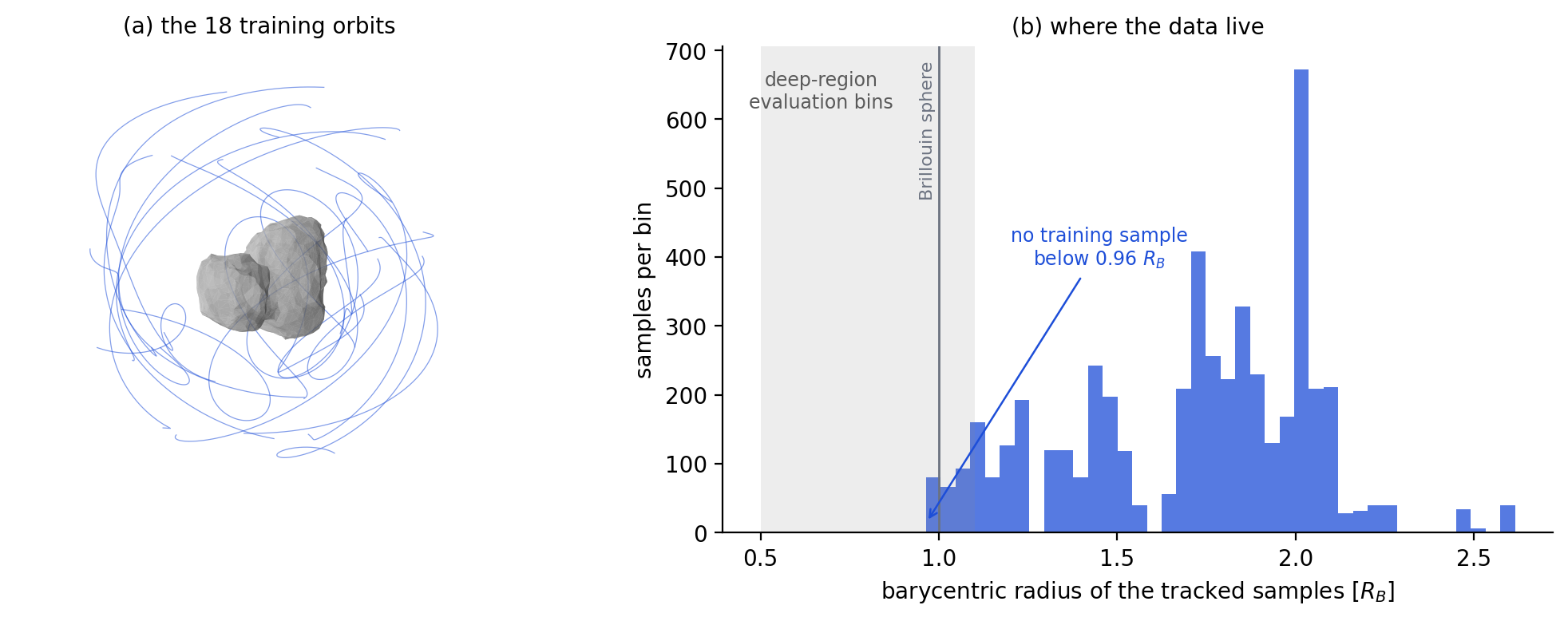}
\caption{Training geometry of the deep-region campaign. (a) The 18
training orbits over the 67P shape, in the body frame. (b) Radial
distribution of the 5040 tracked samples (126 arcs of 40): every
sample lies between 0.96 and $2.6\,\RB$, so the deep-region
evaluation bins below $0.95\,\RB$ (shaded) contain no data at all
--- the deep region is reached by extrapolation, never by
interpolation.}
\label{fig:orbits67p}
\end{figure}

The campaign is the row ``deep region'' of
Table~\ref{tab:campaigns}: homogeneous 67P truth (240 mascons), 18
(not colliding nor escaping) orbits providing 126 arcs, flown in six stages of three
orbits and assimilated via the continual learning procedure discussed in Section~\ref{sec:continual}. The training geometry is shown in
Fig.~\ref{fig:orbits67p}: the orbits envelop the body at flight
altitude, and no tracked sample descends below $0.96\,\RB$, so
everything the models are asked to do in the deep region is
extrapolation.

\begin{figure}[t]
\centering
\includegraphics[width=0.85\linewidth]{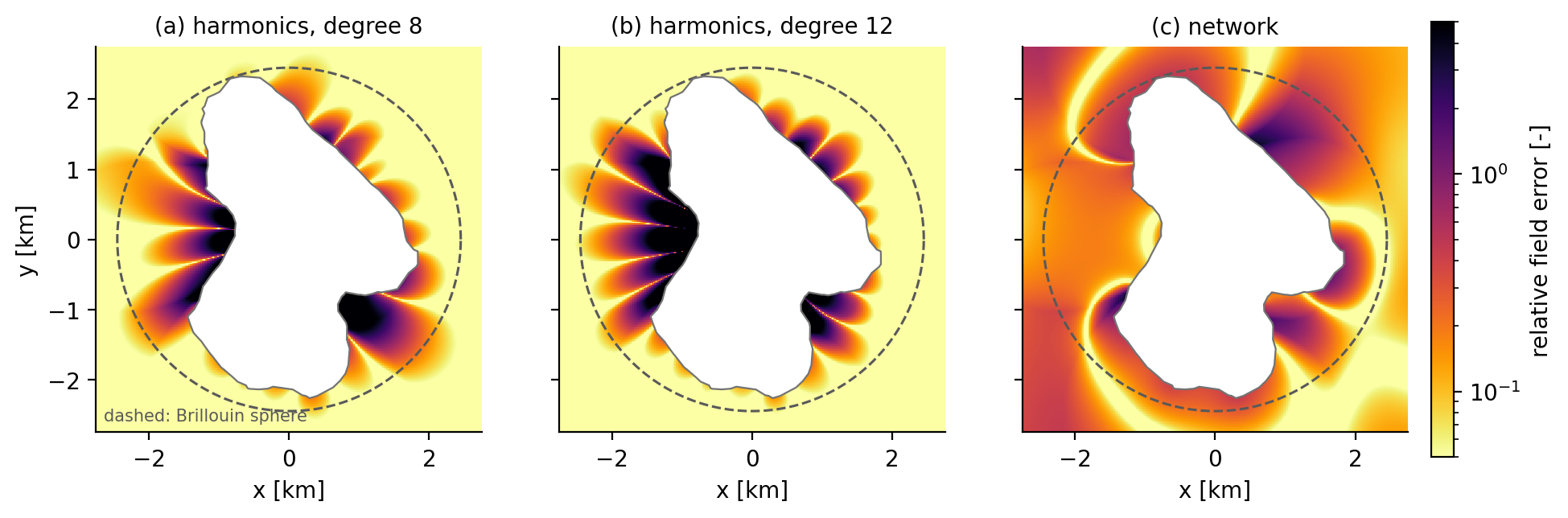}
\caption{Relative field error in the equatorial plane of 67P for the
spherical harmonics models of degree 8 (a) and degree 12 (b), and
for the network (c), on a common logarithmic colour scale; the exact
$z=0$ cross-section of the shape is masked out. Dark lobes adjacent
to the surface are regions where the fitted model is worse than
applying no correction at all.}
\label{fig:maps}
\end{figure}

The
estimators are a 721-parameter network and harmonics of degrees 4
to 12, fitted to the identical arcs. For the harmonics fit, the
accuracy behavior at flight altitude is
the familiar one, with the error decreasing with increasing degree down to
0.012 at degree 8, averaged over shells at the orbit radii
(1.2--1.7\,$\RB$). However, three hundred metres above the surface, inside
the Brillouin sphere over the concave region, the ordering is
inverted: degree 4 yields a relative field error of 0.65, degree 8
of 1.34, i.e.\ worse than applying no correction, and degree 12 of
45. The same coefficients that improve the model at altitude are
amplified by $(\RB/r)^{\ell}$ below it. The network trained on the
identical arcs remains bounded, between 0.29 in the deepest bin and
0.51 at worst across the radial bins, including regions never
visited by the data. Fig.~\ref{fig:maps} maps the same comparison in the equatorial
plane, with the exact cross-section of the shape masked out. For the
harmonics fits, the error exceeds the no-correction level
($\epsilon > 1$) in extended patches adjacent to the surface, the
dark regions of Fig.~\ref{fig:maps} a,b, all lying inside the
Brillouin sphere. Over those patches the median error is 2.7 and
locally exceeds 10. Evaluated over the same patches, the network
error has a median of 0.27, and it exceeds unity only in a thin band
within a few hundred metres of the terrain.

\begin{figure}[t]
\centering
\includegraphics[width=0.78\linewidth]{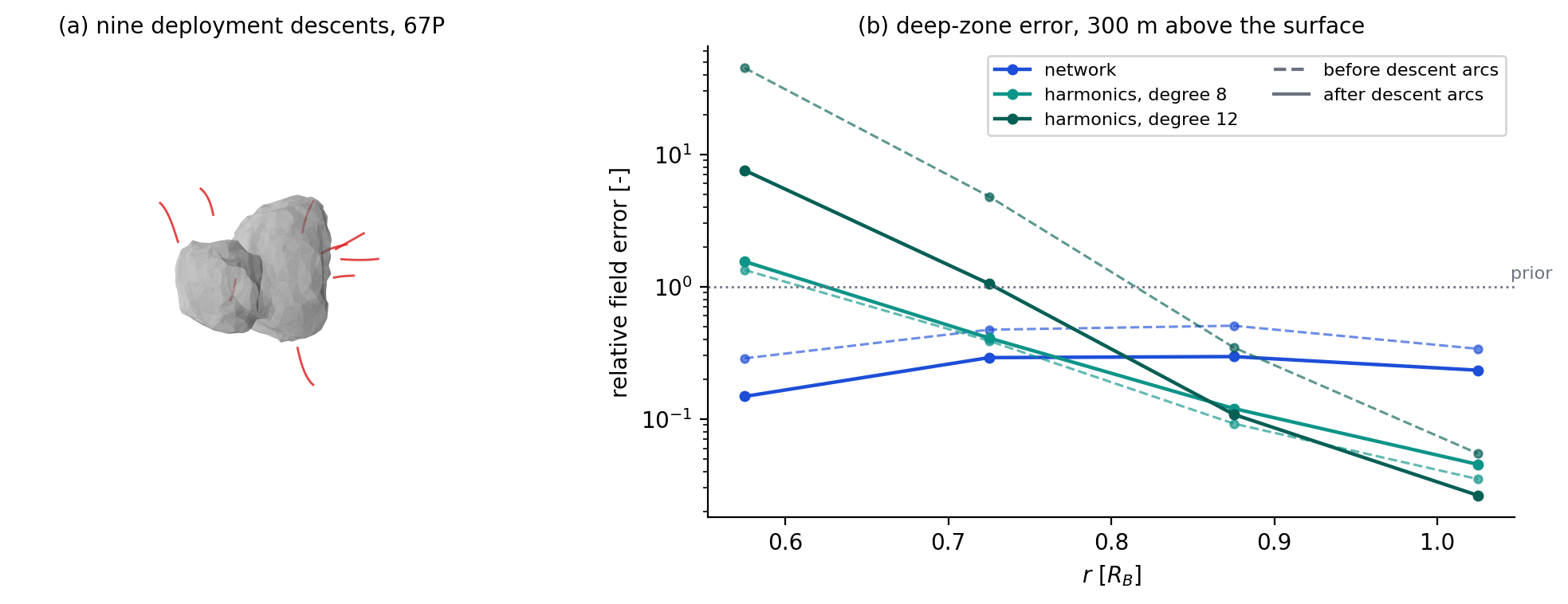}
\caption{Assimilation of descent data on 67P. (a) The nine
deployment descents. (b) Deep-region field error by radial bin,
\SI{300}{m} above the surface, before (dashed) and after (solid)
refitting on the nine descent arcs. The network reaches 0.15 in the
deepest bin; degree 8 lacks the capacity to exploit the same data
and degree 12 diverges on it.}
\label{fig:deepzone}
\end{figure}

The deep region is, moreover, a region whose data the harmonics
cannot assimilate; this is the ``(+9 desc.)'' entry of the same
table row. Adding nine deployment descents to the
orbital arcs and warm-refitting reduces the network's deepest-bin
error from 0.29 to 0.15; refitting degree 8 on the same enlarged
dataset leaves it at 1.5, and degree 12 improves from 45 to 7.6,
still well above the no-correction level (Fig.~\ref{fig:deepzone}).

\subsection{Ballistic Descent Planning}\label{sec:touchdown}

This experiment covers the rows ``descent planning'' and ``extended
campaign'' of Table~\ref{tab:campaigns}.
The
truth is Itokawa with a two-zone interior, the small ``head''
lobe twice as dense as the ``body'', at fixed total mass, a
contrast of the kind suggested by the YORP measurements
of \cite{lowry2014itokawa}. The training
data are orbital arcs only: 16 orbits with radii between 1.2 and
$2.0\,\RB$ and inclinations between 10 and 80$^{\circ}$, with 7 arcs
each, 112 arcs in total, about $2.7\times10^{4}$ scalar measurements,
from which every candidate model (point mass, harmonics of degrees 4
to 12, network) is fitted exactly as in Section~\ref{sec:deep}. The
test is data none of the models have seen:
thirty ballistic descents are released from $1.3\,\RB$ with small
inward manoeuvres (1--3\,\si{cm\,s^{-1}}; median descent duration
about one hour) and propagated to contact under the ground truth,
which yields the actual contact points (Fig.~\ref{fig:orbits}c shows
the descent geometry).
The identical release states are then propagated under each fitted
model, as a mission would do when planning the descent with its
current gravity model, and the miss distance of a descent is the
separation between the predicted and the actual contact point.

\begin{figure}[t]
\centering
\includegraphics[width=0.58\linewidth]{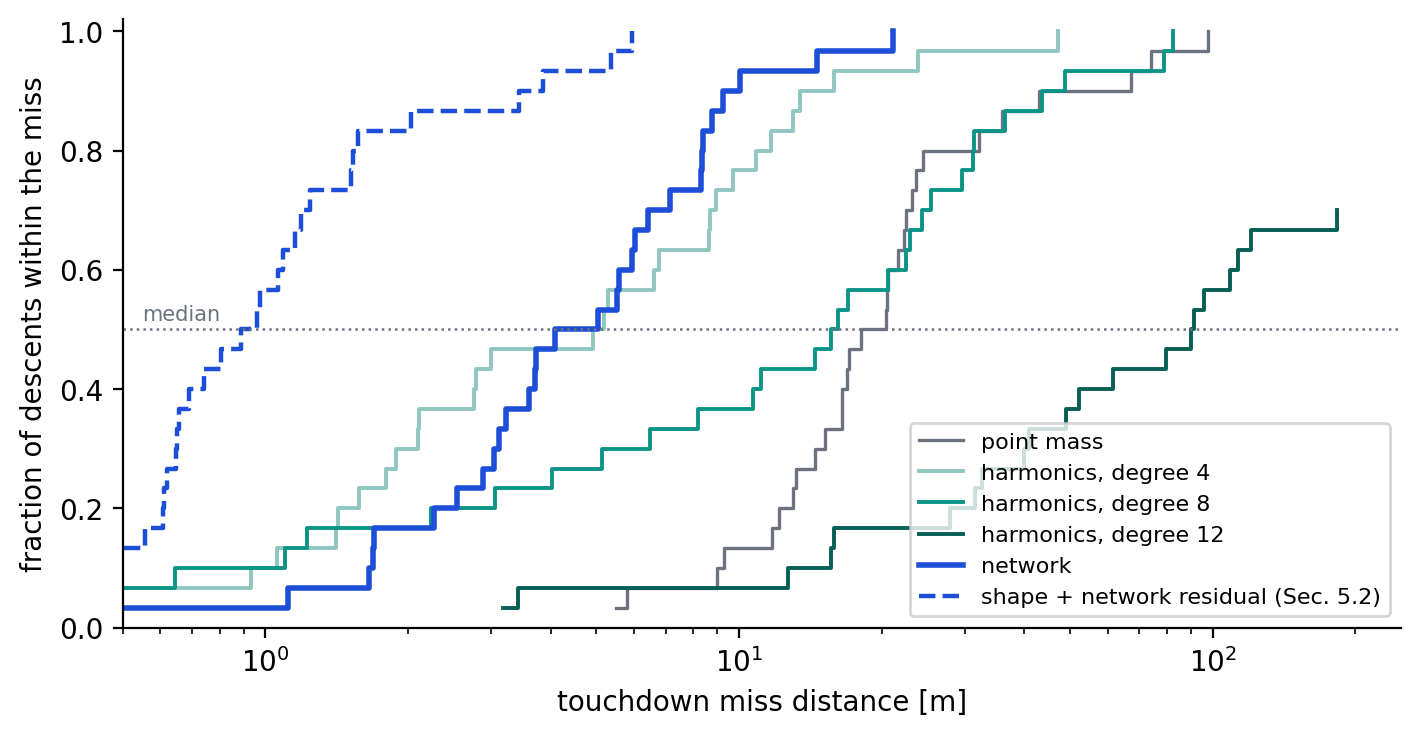}
\caption{The Itokawa descent campaign. Fraction of the 30 descents landing within a given miss
distance, per planning model, all fitted to the same 112 orbital
arcs.}
\label{fig:touchdown}
\end{figure}

With no shape model available, the network is the most accurate
fitted model: \SI{4.6}{m} median and \SI{21.2}{m} worst-case miss,
with all 30 descents reaching the surface. The harmonics fits from
the same arcs yield \SI{5.1}{m} (degree 4), \SI{15.9}{m} (degree 8)
and \SI{48.9}{m} (degree 12) median miss, with 9 of the 30 descents
never reaching the surface under the degree-12 model.
Fig.~\ref{fig:touchdown} shows the full distributions: the median
is where each curve crosses the dotted line, the tails are the upper
right, and models whose curves saturate below one have descents that
never reach the surface. Table~\ref{tab:touchdown} collects the
statistics, and it includes the shape-prior block that will be discussed in
Section~\ref{sec:resid-touchdown}

\begin{table}[t]
\centering\footnotesize
\caption{Touchdown miss distance over the 30 Itokawa descents.
Statistics are computed over the descents that reach the surface
under each model (``landed''); descents that do not are those whose
propagation under the fitted model diverges before contact. Upper
blocks: no shape model (Section~\ref{sec:noshape}); lower blocks:
with the shape prior (Section~\ref{sec:shape}).}
\label{tab:touchdown}
\begin{tabular}{lcccc}
\toprule
planning model & median [m] & 90th pct [m] & worst [m] & landed\\
\midrule
point-mass prior        & 19.3 & 45.4 & 97.9 & 30/30\\
harmonics, degree 4     & 5.1  & 13.7 & 47.1 & 30/30\\
harmonics, degree 8     & 15.9 & 44.2 & 82.7 & 30/30\\
harmonics, degree 12    & 48.9 & 113.0 & 183.4 & 21/30\\
network                 & \textbf{4.6}  & \textbf{9.3}  & \textbf{21.2} & 30/30\\
\midrule
\multicolumn{5}{l}{\footnotesize\emph{extended campaign, 672 arcs:}}\\
harmonics, degree 4     & 2.6  & 7.3  & 49.0 & 30/30\\
harmonics, degree 8     & 1.7  & 10.2 & 14.1 & 29/30\\
network                 & 2.7  & \textbf{6.4}  & 56.2 & 30/30\\
\midrule
shape model (free)      & 3.1  & 8.3  & 29.6 & 30/30\\
shape + harmonics-4 residual & 1.7 & 7.0 & 20.0 & 30/30\\
shape + harmonics-8 residual & 15.4 & 43.8 & 84.3 & 30/30\\
shape + network residual & \textbf{0.9} & \textbf{3.5} & \textbf{6.0} & 30/30\\
\bottomrule
\end{tabular}
\end{table}

A mission is not limited to 112 arcs: proximity operations
produce tracking continuously, and at a few passes per day a
half-year orbital phase yields several hundred arcs, with OSIRIS-REx
and Rosetta each having spent well over a year at their targets. The
row ``extended campaign'' of Table~\ref{tab:campaigns} therefore
repeats the identical experiment, changing only the data budget: 48 orbits
(the same radius and inclination families), each tracked over twice the window with 14 arcs, for
672 arcs and $1.6\times10^{5}$ scalar measurements. Every model fit improves, confirming that the 112-arc errors were partly
estimation-limited: the degree-4 median miss decreases from 5.1 to
\SI{2.6}{m}, the degree-8 one from 15.9 to \SI{1.7}{m}, although one
of the 30 descents still fails to reach the surface under the
degree-8 model, and the network improves from 4.6 to \SI{2.7}{m}
median with the smallest 90th percentile of the campaign
(\SI{6.4}{m}) and all 30 descents landed. At
\SI{30}{m} altitude the network error decreases from 0.52 to 0.35,
while the degree-8 error remains at 3.7, several times worse than
applying no correction. A larger campaign refines the harmonics fit
only inside its convergence domain; it does not mitigate the
divergence in the region where landing takes place.

\subsection{Joint estimation of gravity and solar radiation
pressure}\label{sec:srp}

The force budget of a real mission is never limited to gravity, but other forces often play a role in shaping trajectories around a small body.
Can the same machinery estimate a non-gravitational force
jointly with the field? Second, and operationally
more important, what happens to each representation when a force
that is present in the data is not modelled at all?

The scenario is drawn from past missions: during OSIRIS-REx operations, Bennu was observed to eject centimetre-scale particles
whose tracked trajectories, strongly perturbed by solar radiation
pressure (SRP) due to their large area-to-mass ratios, were studied also
for gravity science
\cite{lauretta2019particles,chesley2020particles}. We reconstruct
the analogous setup, which is the row ``particle swarm'' of
Table~\ref{tab:campaigns}. The truth is the Bennu shape model with
SRP acting on each particle; 28 particles are ejected from the
surface with area-to-mass ratios spanning a factor of about 15, and
tracked over 263 arcs (40 samples each, 39\,s cadence), here the
particles themselves are used as tracking data. On the estimation side, the SRP magnitude of each
particle is unknown: in both the Hamiltonian and the harmonics formulation, this results in 28 coefficients to be estimated jointly with the gravitational ones.  Eq.~\eqref{eq:var} provides exact
sensitivities for all of them through the same variational flow.

\begin{figure}[t]
\centering
\includegraphics[width=0.78\linewidth]{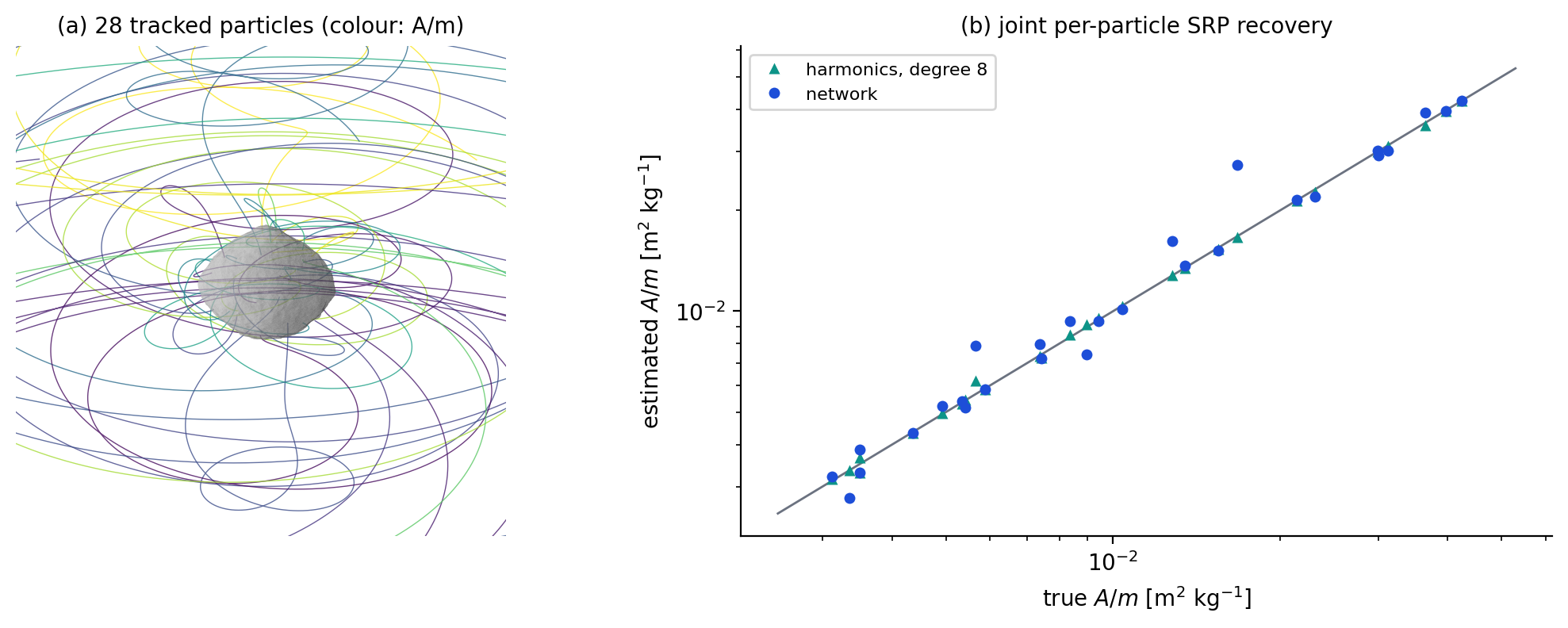}
\caption{The particle-swarm scenario. (a) The 28 tracked particles
over the Bennu shape in the body frame; the colour encodes the true
area-to-mass ratio, which spans a factor of about 15 across the
swarm. (b) Estimated versus true per-particle $A/m$ for the two
joint fits.}
\label{fig:swarm}
\end{figure}

The answer is revealed in Fig.~\ref{fig:swarm}, where it is clear that joint estimation works well for both cases: the per-particle area-to-mass ratio is
recovered with a median error of 0.6\% by the degree-8 harmonics fit
and 3.5\% by the network. The gravity side of the joint fit follows
the pattern already seen in
Section~\ref{sec:deep}: Bennu is nearly
spherical and the particles spend most of their time near apoapsis, and the exterior field is where harmonics shine.

As for what happens to each representation when a force that is present in the data is not modelled at all, we obseve that refitting gravity alone to the same
SRP-active data, the unmodelled force is absorbed into the harmonics
coefficients: the field error at \SI{30}{m} altitude reaches 3.5,
several times worse than applying no model at all. The network under
the identical conditions degrades from 0.33 to 0.48. The $w$ term of Eq.~\eqref{eq:neural} forces the learned potential
to decay as $r_1^{-3}$, and makes it more difficult to absorb forces that do not fulfill that behavior.

\section{Results with the Imaged Shape as a Prior}\label{sec:shape}

Once imaging has delivered the shape, the constant-density shape
model constitutes strong prior knowledge (for a homogeneous body it
coincides with the truth), and we leverage the same machinery, but moving the prior knowledge into $V_{\mathrm{known}}$ of Eq.~\eqref{eq:ham},
leaving to $\mathcal{N}$ only the residual generated by the
interior. That a model built from the imaged geometry outperforms
models estimated from arcs alone is expected, and we confirm it
below. The questions of interest are what tracking adds on top of
the shape model, and which representation exploits it the most. The section
covers the remaining rows of Table~\ref{tab:campaigns}: the
heterogeneous-interior bodies
(Section~\ref{sec:architecture}, row ``interiors''), the descent
campaign repeated with the prior in place
(Section~\ref{sec:resid-touchdown}, row ``descent, residual''), and
the buried anomalies (Section~\ref{sec:nuggets}, row
``anomalies'').

The shape enters the Hamiltonian in the same representation as the
ground truth, as a mascon set. The imaged shape, filled with uniform
density, is reduced to $N_s$ point masses $\{\bm q_j, \hat m_j\}$:
the positions are the centroids of a $k$-means tessellation of the
shape volume, and the masses are then fit by
non-negative least squares so that the reduced set reproduces the
exterior potential of the full-resolution constant-density model,
sampled between 1.02 and $1.8\,\RB$, with $\sum_j \hat m_j = 1$ and
the origin at the centre of mass. This step is a compression of the
imaged shape model, not part of the estimation: after this fitting is performed, the mascons positions and masses never change
afterwards. The known part of
Eq.~\eqref{eq:ham} then becomes
\begin{equation}
\frac{1}{r} + V_{\mathrm{known}}(\bm q)
 = \sum_{j=1}^{N_s} \frac{\hat m_j}{\lVert \bm q - \bm q_j \rVert}
 \equiv V_{\mathrm{shape}}(\bm q).
\label{eq:vshape}
\end{equation}
Hence, the central term and the shape-model perturbation are carried
jointly as the mascon sum, compiled symbolically into the equations
of motion like every other term. The target of the estimated correction changes
accordingly: $\mathcal{N}$ no longer represents the whole perturbing
field but only the interior residual,
\begin{equation}
\mathcal{N}(\bm q;\bm\theta) \;\approx\; \delta V_{\mathrm{int}}(\bm q)
 = \sum_{i=1}^{N_m} \frac{\mu_{m,i}}{r_{m,i}} - V_{\mathrm{shape}}(\bm q),
\label{eq:residtarget}
\end{equation}
the difference between the true mascon potential of
Eq.~\eqref{eq:truth} and that of the same shape at constant density.
The workflow is thus: imaging fixes
$V_{\mathrm{shape}}$, tracking estimates $\bm\theta$, and nothing
else is refitted; the estimated object is a scalar field, not a
set of movable masses. Two properties of this residual follow from
the construction. Truth and prior carry the same total mass, so
$\delta V_{\mathrm{int}}$ has no monopole; and when both are centred
on the centre of mass, as in the experiments of
Sections~\ref{sec:architecture} and~\ref{sec:resid-touchdown}, it
has no dipole either, so its leading multipole is a quadrupole: this means that the same $w$ term of Eq.~\eqref{eq:neural} is still appropriate here and hence remains unchanged.  The harmonics
baseline occupies the same role in the same way, as an expansion of
Eq.~\eqref{eq:sh} estimated on top of the shape prior. Nothing in
the estimation changes: the solvers of Section~\ref{sec:solvers} act
on the same arc residuals through the same variational flow machinery, and
moving the shape into the known part amounts to one recompilation of
the dynamics, not to a modification of the estimator.

\subsection{The Interior Residual and the Choice of Architecture}
\label{sec:architecture}
The residual that the shape model cannot explain is not small. On
the bodies with heterogeneous interiors of the row ``interiors'' of
Table~\ref{tab:campaigns} (an Itokawa head/tail contrast, and a
body with an off-centre void amounting to 18\% of the volume), the
constant-density assumption misses 15--21\% of the perturbing
field already for the moderate, YORP-motivated factor-two Itokawa
contrast (the same truth as in Section~\ref{sec:touchdown}).
Imaging constrains the shape; the interior remains a free
variable that only tracking can determine \cite{lowry2014itokawa}.

\begin{figure}[t]
\centering
\includegraphics[width=0.78\linewidth]{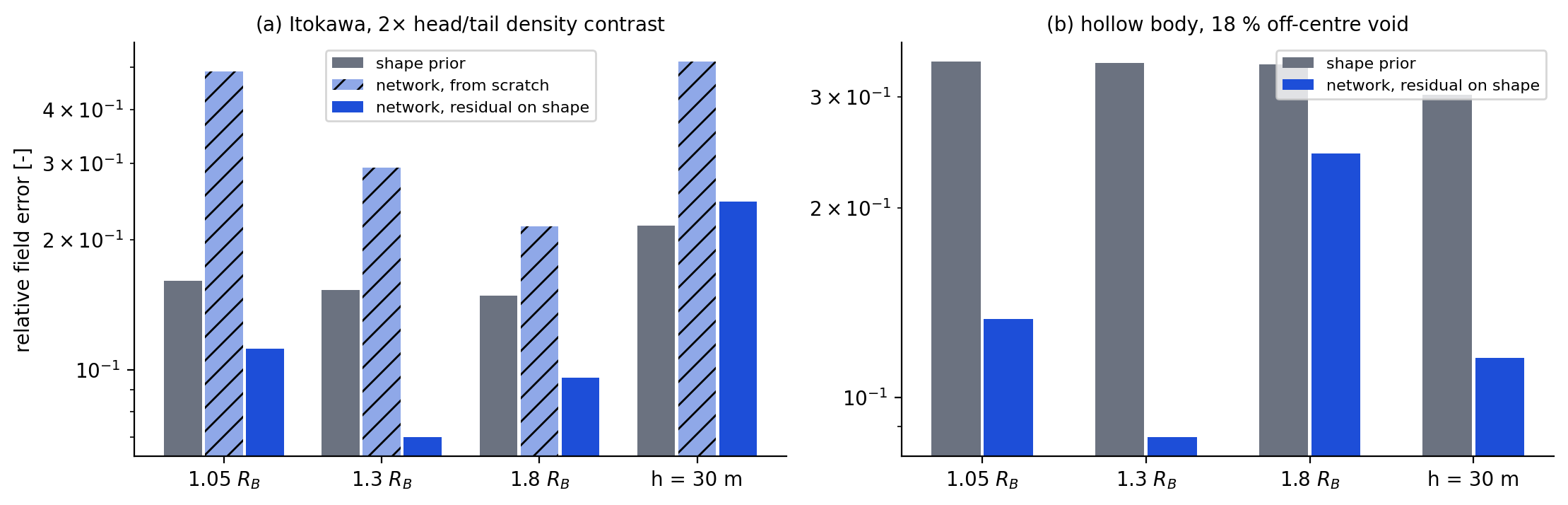}
\caption{The residual architecture. (a) Itokawa with a 2$\times$
head/tail density contrast: the same network fitted from scratch and
as a residual on the shape prior. (b) The hollow body: the residual
reduces the shape prior error by more than a factor of two in every
bin.}
\label{fig:residual}
\end{figure}

Fitting the network from
scratch on these bodies, i.e.\ ignoring the available shape model,
wastes the information it carries: the from-scratch fit reaches a
relative field error of 0.489 at $1.05\,\RB$, where the constant density
shape model alone stands at 0.160. The same network, with the same
data and budget, trained as a residual on the shape model, improves
on the prior at every altitude and on its own from-scratch version
by a factor 4.4 (Fig.~\ref{fig:residual}). This is also the intended
reading of Section~\ref{sec:noshape}: the from-scratch configuration
is for the mission phase in which nothing better exists, and every
piece of knowledge should be moved into $V_{\mathrm{known}}$ as soon
as it becomes available.

\subsection{Descent planning with the residual model}
\label{sec:resid-touchdown}

Repeating the descent campaign of Section~\ref{sec:touchdown} with
the shape prior in place (row ``descent, residual'' of
Table~\ref{tab:campaigns}, lower block of
Table~\ref{tab:touchdown}, and Fig.~\ref{fig:touchdown}), the constant density shape model plans to
\SI{3.1}{m} median miss, better than every model estimated from arcs
alone, as expected. Adding a harmonics residual improves the
planning at degree 4 (\SI{1.7}{m}) and degrades it at degree 8
(\SI{15.4}{m}); adding the network residual results in \SI{0.9}{m}
median and \SI{6.0}{m} worst-case miss with no failures, a factor
3.4 better than the shape model alone and the best result of the
campaign by a factor of about two. 

\subsection{Localized density anomalies}\label{sec:nuggets}

The interior structure of rubble piles is expected to be localized:
discrete blocks, and at the largest scale contrasts such as the
1.75 versus 2.85 g\,cm$^{-3}$ inferred for Itokawa's two lobes
\cite{lowry2014itokawa}. A global basis is poorly suited for localized masses. Resolving a
feature of angular size $\theta$ requires degrees from
$L\sim\pi/\theta$ upwards, and representing a compact peak while
cancelling its sidelobes requires substantially higher degree still.
Those high degrees, however, are the ones tracking can barely see:
at an observation radius $r$ above the Brillouin sphere, the
degree-$\ell$ content of the potential carries the factor
$(\RB/r)^{\ell+1}$, which decays exponentially with $\ell$, each
additional degree is damped further, so
that from typical arc radii the high-degree signal lies below the
measurement noise, and no amount of tracking at that altitude can
constrain the corresponding coefficients. A localised unknown thus
puts the global basis in a very complex situation:
the representation demands high degrees, and the estimation cannot
supply them. The network, instead, is more adaptive and should be able to represent
localised features with a few units, placing capacity only where the
data requires it.

The experiment is constructed to isolate this effect. We use the
Bennu shape, which is nearly spherical, so that every evaluation
point is exterior, and no Brillouin-sphere effect is present, and add
four dense anomalies buried at 0.85 of the local surface radius,
each amounting to 1.5\% of the total mass, with the background mass
rescaled so that the total is unchanged
(Fig.~\ref{fig:orbits}a shows the training orbits). The residual that the shape model cannot explain amounts to
76--84\% of the total perturbing signal on the exterior shells of
Table~\ref{tab:nuggets} (row ``shape prior alone'').  The campaign is the row ``anomalies''
of Table~\ref{tab:campaigns}: sixteen orbits between 1.08
and $1.7\,\RB$ provide 112 arcs. Every estimator is a residual on
the shape model, and the classical side includes degrees 4 to 16 as well
as Kaula-regularised fits at degrees 8 and 16, with $\alpha$ scanned
over a range and the best value retained.

\begin{figure}[t]
\centering
\includegraphics[width=0.85\linewidth]{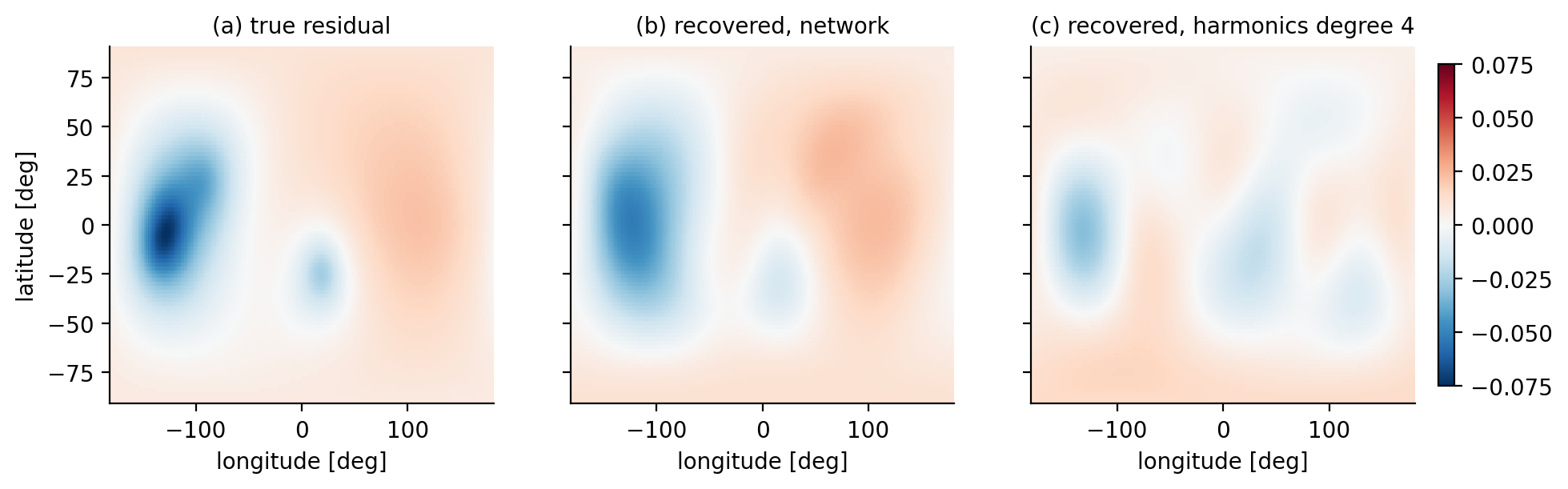}
\caption{The body with buried anomalies, $1.05\,\RB$ shell, common
colour scale. (a) The residual that the shape model cannot explain.
(b) The recovery by the network, concentrated where the anomalies
are located. (c) The recovery by the best harmonics fit (degree 4).}
\label{fig:nuggetmaps}
\end{figure}

Table~\ref{tab:nuggets} reports the
outcome. On a nearly spherical body with all evaluation points exterior:
degree 4 improves moderately on the prior (0.63 against 0.76),
degree 8 is worse than applying no correction, degree 16 is worse by
a factor of 45, and the best-$\alpha$ Kaula fits only return to the
prior error. The network residual reaches
0.17--0.19 on every shell, a factor 3.5 better than the best
harmonics option over all degrees and regularizations; given that
the prior explains only a quarter of the signal, the reduction from
0.76 to 0.18 is attributable to the network. The true residual, together with the one recovered by the network and the one recovered by the best harmonics fit, are shown in Fig.~\ref{fig:nuggetmaps}, confirming the superior capability of the network to accurately represent the anomalies.

\begin{table}[t]
\centering\footnotesize
\caption{Relative field error on all-exterior shells of the body
with buried anomalies. Every model is a residual on the
constant-density shape prior.}
\label{tab:nuggets}
\begin{tabular}{lcccc}
\toprule
model & $1.03\,\RB$ & $1.05\,\RB$ & $1.1\,\RB$ & $1.2\,\RB$\\
\midrule
shape prior alone        & 0.758 & 0.767 & 0.792 & 0.844\\
+ harmonics, degree 4    & 0.629 & 0.639 & 0.667 & 0.725\\
+ harmonics, degree 8    & 1.177 & 1.141 & 1.071 & 0.996\\
+ degree 8, Kaula (best $\alpha$)  & 1.060 & 1.033 & 0.983 & 0.936\\
+ harmonics, degree 12   & 3.846 & 3.490 & 2.801 & 1.985\\
+ harmonics, degree 16   & 51.82 & 44.90 & 32.19 & 18.27\\
+ degree 16, Kaula (best $\alpha$) & 0.953 & 0.929 & 0.886 & 0.855\\
+ network residual       & \textbf{0.189} & \textbf{0.181} & \textbf{0.170} & \textbf{0.174}\\
\bottomrule
\end{tabular}
\end{table}

\section{Conclusions}\label{sec:conclusions}

We have presented a method that learns the unknown part of the
dynamics around a small body as a neural Hamiltonian, estimated from
tracking arcs through the exact variational equations of a Taylor
integrator, and we have assessed it against the spherical harmonics
pipeline estimated from identical data through identical numerics.

For smooth exterior gravity observed from altitude, i.e.\ the radio
science regime that occupies most of a mission, the classical
expansion is the more accurate estimator by a wide margin. However,
the phases of a small-body mission that place the highest demands on
the gravity model, namely descent, landing, sampling and proximity
operations over irregular terrain, are also the phases in which the
classical expansion fails for structural reasons: it diverges inside
the Brillouin sphere, it cannot observe localised interior structure
from finite tracking, and it absorbs force model errors into its
coefficients. The proposed method serves these regimes from tracking
alone, before any shape model exists, and incorporates the shape
model as known physics as soon as imaging provides it: from tracking
alone it is the most accurate model at touchdown and the only one
able to assimilate descent data. With the shape as prior it reaches
metre-level descent planning and recovers the effect of buried
density anomalies inaccessible to any harmonics degree or
regularisation. It also degrades mildly, rather than
catastrophically, under force mis-modelling. The two representations fail in
disjoint regimes, and are best flown together across the phases of a
small-body mission.

Finally, the estimation procedure is continual by construction: warm refits
assimilate each batch of tracking as the campaign grows, and the
same information matrix that is used by the solvers can be leveraged to score candidate
orbits before they are flown.

{\small
\bibliographystyle{unsrt}
\bibliography{references}}

\end{document}